\documentclass[manuscript]{aastex}
\usepackage{natbib}
\usepackage{amsmath}

\newcommand{\be}{\begin{displaymath}}
\newcommand{\ee}{\end{displaymath}}

\def\lsim{\hbox{\rlap{\raise 0.425ex\hbox{$<$}}\lower 0.65ex\hbox{$\sim$}}}
\def\gsim{\hbox{\rlap{\raise 0.425ex\hbox{$>$}}\lower 0.65ex\hbox{$\sim$}}}

\shorttitle{Long term monitoring of mode switching for PSR B0329+54}
\shortauthors{Chen et al.}

\begin{document}

\title{Long term monitoring of mode switching for PSR B0329+54}

\def\cfa{1}
\def\berk{2}
\def\clay{3}
\def\chikav{4}
\def\chiast{5}
\def\cape{6}
\def\saao{7}
\def\fermi{8}
\def\nd{9}
\def\rut{10}
\def\tok{11}
\def\port{12}
\def\jhu{13}
\def\stsci{14}
\def\penn{15}
\def\psu{16}
\def\stock{17}
\author{J. L. Chen\altaffilmark{1,2}, H. G. Wang\altaffilmark{3,5} N. Wang\altaffilmark{1}, A. Lyne\altaffilmark{4}, Z. Y. Liu\altaffilmark{1}, A.
Jessner\altaffilmark{5}, J. P. Yuan\altaffilmark{1}, M.
Kramer\altaffilmark{5}}
\email{hgwang@gzhu.edu.cn}
\altaffiltext{1}{Xinjiang Astronomical Observatory, 150, Science-1
Street, Urumqi, Xinjiang, 830011, China} \altaffiltext{2}{Graduate
University of the Chinese
   Academy of Sciences, 19A Yuquan road, Beijing, 100049, China.}
\altaffiltext{3}{Center for Astrophysics, Guangzhou University,
Guangzhou 510006, China; } \altaffiltext{4}{University of
Manchester, Jodrell Bank, England.} \altaffiltext{5}{Max-Planck
Institute for Radio Astronomy, Bonn 53121, Germany.}

\begin{abstract}
The mode switching phenomenon of PSR B0329+54 is investigated based
on the long-term monitoring from September 2003 to April 2009 made
with the Urumqi 25m radio telescope at 1540~MHz. At that frequency,
the change of relative intensity between the leading and trailing
components is the predominant feature of mode switching. The
intensity ratios between the leading and trailing components are
measured for the individual profiles averaged over a few minutes. It is
found that the ratios follow normal distributions, where the
abnormal mode has a wider typical width than the normal mode,
indicating that the abnormal mode is less stable than the normal mode.
Our data show that 84.9\% of the time for PSR B0329+54 was in the normal
mode and 15.1\% was in the abnormal mode. From the two passages of
eight-day quasi-continuous observations in 2004, and supplemented by
the daily data observed with 15~m telescope at 610~MHz at Jodrell
Bank Observatory, the intrinsic distributions of mode timescales are
constrained with the Bayesian inference method. It is found that the gamma
distribution with the shape parameter slightly smaller than 1 is
favored over the normal, lognormal and Pareto distributions. The
optimal scale parameters of the gamma distribution is 31.5 minutes for
the abnormal mode and 154 minutes for the normal mode. The shape
parameters have very similar values, i.e. 0.75$^{+0.22}_{-0.17}$ for
the normal mode and 0.84$^{+0.28}_{-0.22}$ for the abnormal mode, indicating
the physical mechanisms in both modes may be the same. No long-term
modulation of the relative intensity ratios was found for both the
modes, suggesting that the mode switching was stable.
The intrinsic timescale distributions, for the first time
constrained for this pulsar, provide valuable information to
understand the physics of mode switching.

\end{abstract}





\section{Introduction}

It is well known that most radio pulsars have very stable average
pulse profiles, however, some pulsars are exceptional, showing two
or more patterns of average profiles, which is called the mode
switching (or mode changing) phenomenon. The pattern lasting for a
longer time is usually regarded as the normal mode while the others as
abnormal modes. Since the first discovery of mode switching in PSR
B1237+25 (Backer 1970), this phenomenon has been seen in a few
dozens of pulsars. Many of them are the pulsars which have complex pulse
profiles with both the ``core'' and ``conal'' components (Rankin 1983,
1986). Some authors pointed out that the normal mode is usually distinct from the abnormal modes in the polarization and sub-pulse drifting properties (Rankin 1988, Suleymanova \& Izvekova 1998).

PSR B0329+54 is an excellent candidate for studying mode switching
because of its complex integrated profiles, high
intensity, and frequent occurrence of mode switching. It was found
that at 408~MHz two relative intensities, i.e. the ratio between the leading
outer and the core components, and the ratio between the trailing outer
and the core components, changed simultaneously during mode
switches, and the abnormal mode persisted for several tens of
minutes (Lyne 1971). Observations at 2.7~GHz and 14.8~GHz showed
that the abnormal pulse profiles at higher frequencies are different
from that at 408~MHz (Hesse 1973, Bartel 1978).

Bartel et al. (1982) summarized the observations for PSR B0329+54
over a wide range of frequencies from 410~MHz to 14.8~GHz and concluded
that the pulse phase, spectral and polarization properties shew a
remarkable variation when mode switching occurred. During a mode
switch, the total pulse width becomes narrower because the profile
components change their phases. The spectra of components steepen or
flatten when the pulsar switches to the abnormal mode. All the
polarization features, especially the linear polarization position
angle, are affected by mode changing. It was also found that the
modes switched simultaneously at 1.4 and 9.0~GHz.

Further progress was made by Bartel et al. (1982) through a 2.3 day monitoring for PSR
B0329+54. They concluded that the duration
of the abnormal mode lasted from a couple of pulse periods to several hours
and the switching time could occur within one pulse period. The
authors divided the abnormal mode into three types at 1.4~GHz for
the first time. The observations were extended to a lower frequency
111.4~MHz by Suleimanova \& Pugachev (2002), who revealed that in
half the cases the amplitudes of two outer components varied
simultaneously during mode switches, while in the remaining cases
only the amplitude of the trailing outer component changed
significantly.

In the first polarimetry observations at 10.5~GHz made by Xilouris et al. (1995), it was found that the occurrence rate of abnormal mode is consistent with that at 1.4~GHz, while the polarization properties are very different from those at 1.7~GHz (Bartel et al. 1978). The mode
switching of PSR B0329+54 was clearly visible at 32~GHz (Kramer et al. 1996), where one component was seen in the normal mode, while two components were visible in the abnormal mode. The switching was detected at hitherto the highest
frequency up to 43~GHz (Kramer et al. 1997), which verified the observation by Bartel et al. (1982) that each mode can
last as long as 65 minutes, but did not confirm their observation that the abnormal-mode pulse profile is more frequency dependent.

The previous observations usually lasted not more than a few hours, hence it
is impossible to obtain a full knowledge on the moding timescales.
In this paper, we investigated the temporal properties of moding
events of PSR B0329+54 by analyzing the data from 2003 to 2009
observed with the 25m telescope at Urumqi Astronomical Observatory (UAO) at 1.5 GHz, including two passages
of eight-day quasi-continuous observations in March 2004, and the data obtained with the 15m telescope at Jodrell Bank Observatory (JBO) at 610~MHz. Details of the
observations are described in \S 2. Methods of data reduction and
results are presented in \S 3. Conclusions and discussions are made
in \S 4.


\section{Observations}
The data were collected from a long-term timing project carried on
with the Nanshan 25m telescope at UAO. The de-dispersion was
provided by a $2 \times128\times$2.5~MHz filter bank system {\rm in}
a total bandwidth of 320~MHz. More detailed descriptions of the
observing systems are referred to Wang et al. (2001). To obtain enough signal to noise ratio, about 80 pulse periods ($\sim$60s)
were averaged and a time resolution of 256 bins per pulse period was
used in all observations.
\begin{figure*}[h!!!]
\centering
   \includegraphics[width=10.0cm, angle=0]{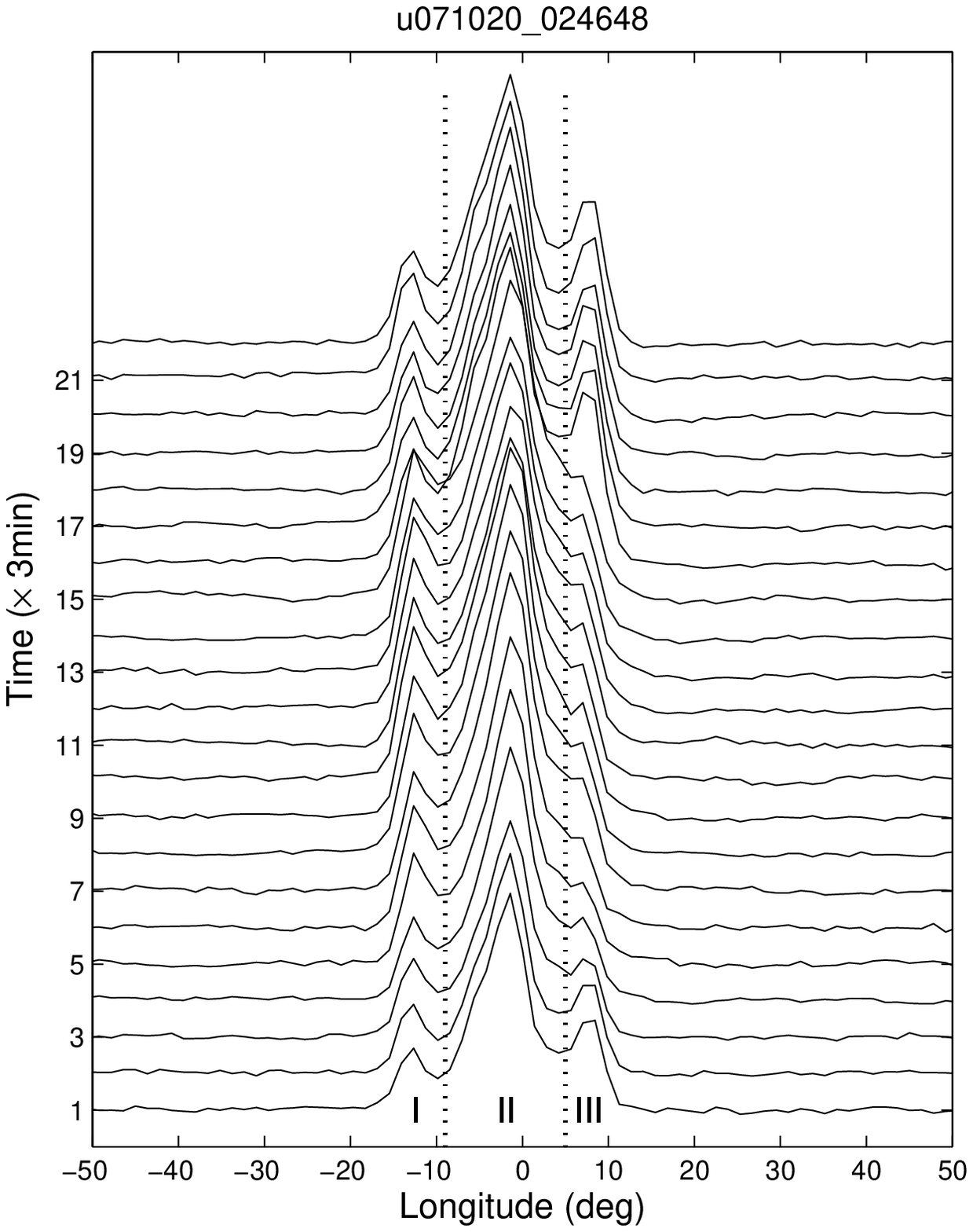}
   \caption{A sequence of pulse profiles observed in 2007 October 20 which contained both the normal and abnormal modes, each averaged over 3 minutes. The 4th to the 15th integrated profiles show switches to the abnormal mode. The dotted lines are the inner
boundaries that divide the profiles into three components, I, II and
III.}
   \label{Fig: pulsecomponent}
   \end{figure*}

Two kinds of data in the total 521-hour observations from 2003 to
2009 were used in this work. The
first kind consists of two long passages of quasi-continuous
observations between 2004 March 12 and March 31. The observation
lasted for $182$ hr in the first 8-d passage, and then $178$ hr in the
second passage after an interruption of 2.3 days. They are very
useful for studying the properties of moding timescales. The second
kind consists of the other 90 separated observations from 2003 to 2009,
which lasted for at least 0.5 hour and mostly longer than 2 hours (see Table
\ref{table:MJD}). These data are used to check the long-term variability of mode
switching.

\begin{table}
\footnotesize
\begin{center}
\caption{The selected 90 separated observations with the 25m
telescope at UAO. }
 \label{table:MJD}
  \begin{tabular}{ccccccccc}
  \hline
 Obs. Date     & MJD      & $t_{\rm tot}$ &Obs. Date  & MJD      & $t_{\rm tot}$ & Obs. Date     & MJD      & $t_{\rm tot}$  \\
(yymmdd)&& (hr)&(yymmdd)&& (hr)&(yymmdd)&& (hr)\\
\hline
030924& 52906.1 & 2.1 &  061230 & 54099.7 & 3.0 &    080630& 54647.7& 2.0   \\
031207& 52980.8 & 2.0 &  070128 & 54128.6 & 0.7 &    080705& 54652.8& 2.0   \\
031210& 52983.6 & 2.0 &  070226 & 54157.7 & 1.5 &    080712& 54659.3& 2.0   \\
031220& 52993.7 & 1.1 &  070320 & 54179.5 & 3.0 &    080716& 54663.1& 2.0   \\
040104& 53008.6 & 2.0 &  070409 & 54199.6 & 0.6 &    080726& 54673.1& 2.0   \\
040117& 53021.8 & 2.0 &  070625 & 54276.3 & 2.2 &    080727& 54674.3& 2.0   \\
040207& 53042.4 & 2.0 &  070714 & 54295.3 & 2.4 &    080807& 54685.9& 1.9   \\
040216& 53051.6 & 1.8 &  071020 & 54393.1 & 3.0 &    080815& 54693.0& 2.0   \\
040522& 53147.4 & 1.7 &  071216 & 54450.8 & 3.0 &    080827& 54705.1& 2.0   \\
040523& 53148.3 & 1.2 &  071217 & 54451.8 & 3.0 &    080908& 54717.7& 2.0   \\
051209& 53713.7 & 0.9 &  071228 & 54462.3 & 0.6 &    080920& 54729.2& 2.0   \\
051220& 53724.0 & 1.0 &  071228 & 54462.3 & 2.0 &    080927& 54736.6& 2.0   \\
051220& 53724.4 & 1.0 &  080107 & 54472.4 & 2.0 &    081007& 54746.0& 2.0   \\
051225& 53729.5 & 1.0 &  080116 & 54481.7 & 2.0 &    081010& 54749.7& 2.0   \\
060116& 53751.7 & 2.0 &  080128 & 54493.6 & 2.0 &    081023& 54762.6& 2.0   \\
060126& 53761.6 & 3.0 &  080130 & 54495.0& 2.0  &    081102& 54772.5& 0.9   \\
060209& 53775.6 & 1.4 &  080222 & 54518.4& 2.0  &    081102& 54772.6& 2.0   \\
060209& 53775.7 & 1.5 &  080302 & 54527.2& 2.0  &    081115& 54785.1& 2.0   \\
060221& 53787.2 & 1.0 &  080304 & 54529.7& 2.0  &    081130& 54800.7& 2.0   \\
060310& 53804.2 & 2.0 &  080330 & 54555.2& 1.7  &    081208& 54808.7& 2.0   \\
060314& 53808.4 & 0.8 &  080408 & 54564.5& 2.0  &    081219& 54819.2& 2.0   \\
060331& 53825.9 & 1.6 &  080409 & 54565.4& 2.0  &    081229& 54829.8& 2.0   \\
060402& 53827.4 & 1.0 &  080416 & 54572.0& 2.0  &    081230& 54830.9& 1.6   \\
060412& 53837.2 & 1.0 &  080427 & 54583.5& 2.0  &    090106& 54837.9& 1.8   \\
060424& 53849.1 & 2.7 &  080510 & 54596.4& 2.0  &    090120& 54851.1& 2.0   \\
061126& 54065.8 & 0.7 &  080519 & 54605.4& 2.0  &    090212& 54874.3& 2.0   \\
061207& 54076.8 & 0.7 &  080614 & 54631.3& 2.0  &    090219& 54881.4& 2.0   \\
061209& 54078.8 & 0.7 &  080615 & 54632.4& 2.0  &    090303& 54893.4& 1.5   \\
061221& 54090.8 & 0.7 &  080617 & 54634.3& 1.8  &    090318& 54908.6& 2.0   \\
061230& 54099.6 & 1.1 &  080630 & 54647.1& 1.9  &    090405& 54926.5& 2.0   \\
\hline
 \end{tabular}
 \end{center}
\begin{flushleft}
$t_{\rm tot}$ is the total time of each observation.
\end{flushleft}
\end{table}

\begin{figure*}[h!!!]
   \includegraphics[width=15.0cm, angle=0]{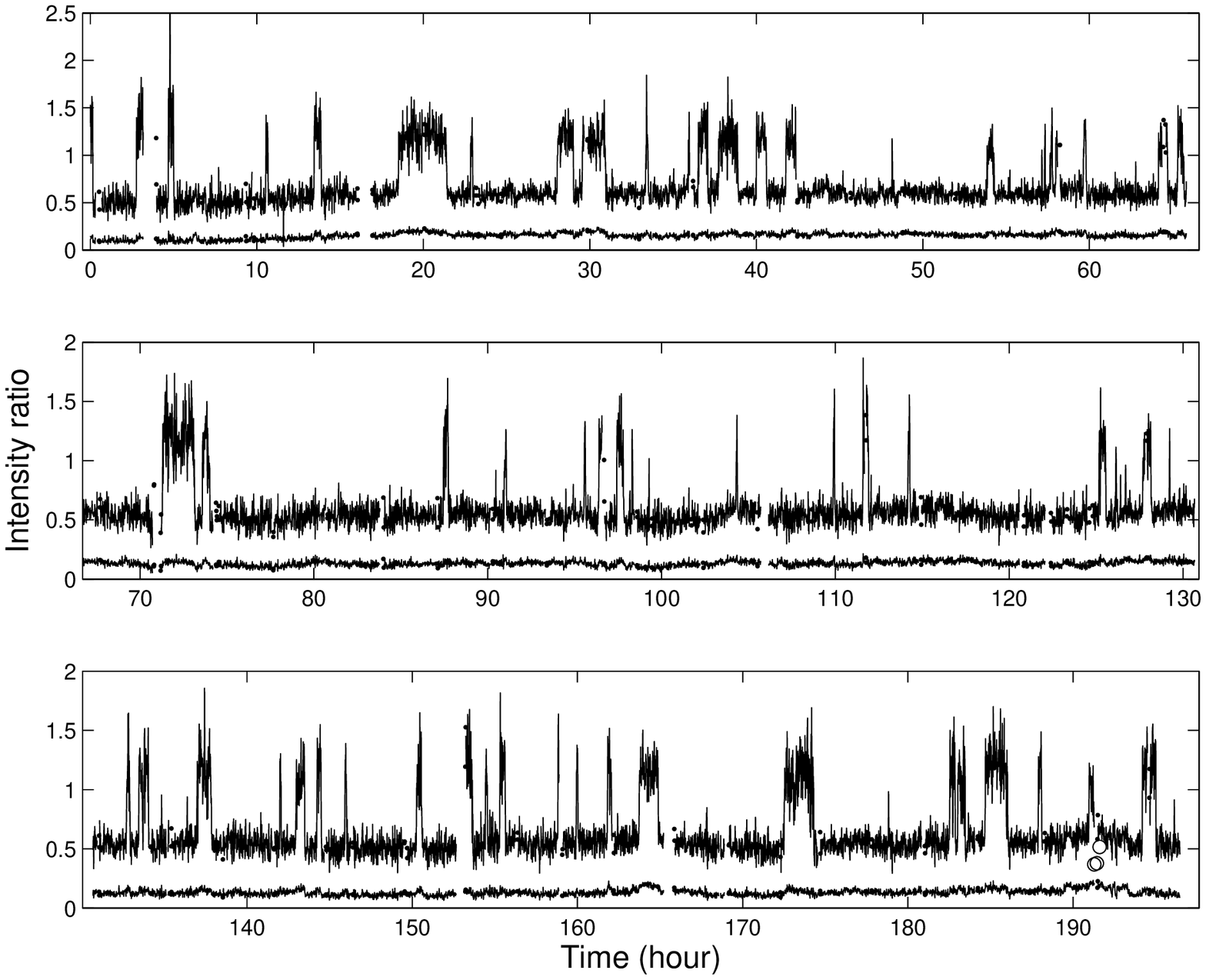}
   \caption{Time sequence of $R$ and  $R^\prime$ for the first 8-day quasi-continuous observation from 2004 March 12 to
March 20. The upper curve in each panel is the intensity ratio $R$
between the components I and III, the lower curve is the ratio
$R^\prime$ between the components 1 and II. The integration time for
individual profiles is 1 minute. The circles, of which the values are the ratios of $R$ at 610~MHz, represent the gap which
are filled by the data of Jodrell Bank Observatory. }
   \label{Fig:eightday-former}
   \end{figure*}

\begin{figure*}[h!!!]
   \includegraphics[width=15.0cm, angle=0]{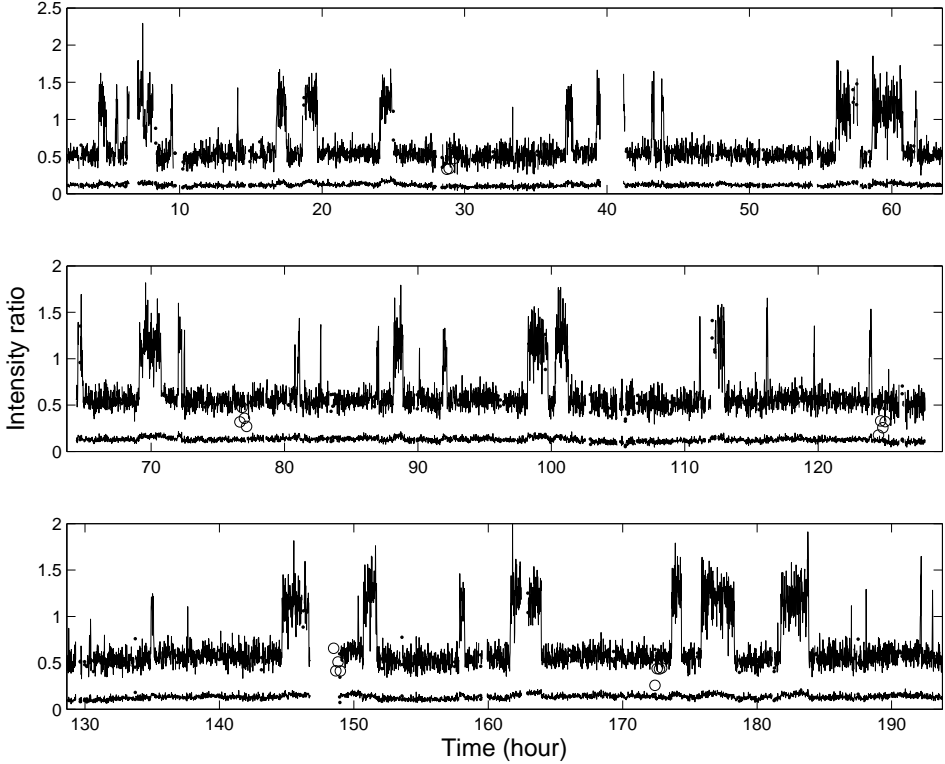}
   \caption{The time sequence of $R$ and $R^\prime$ for the second 8-day quasi-continuous observation from 2004 March 23 to March 31.}
   \label{Fig:eightday-later}
   \end{figure*}

To fill the gaps between the sub-windows of quasi-continuous
observations in 2004 (see more details in \S 3.2), we applied the
monitoring data at 610 MHz for PSR B0329+54 obtained with the 15~m telescope at
JBO. The observations
were made 3 to 5 times nearly everyday since 1990. Each of them
lasted for about 10 minutes and yielded an integrated profile. Only
the integrated profiles are available now. These observations are
helpful to verify the modes at 1540 MHz under the scenario that mode
switches happen simultaneously at different frequencies, which is
supported by previous observations (Bartel et al. 1982). Six gaps in the
UAO data are filled by the JBO data, as shown by the circles in
Figs.\ref{Fig:eightday-former} and \ref{Fig:eightday-later}.

\begin{figure*}[h!!!]
   \includegraphics[width=18.0cm, angle=0]{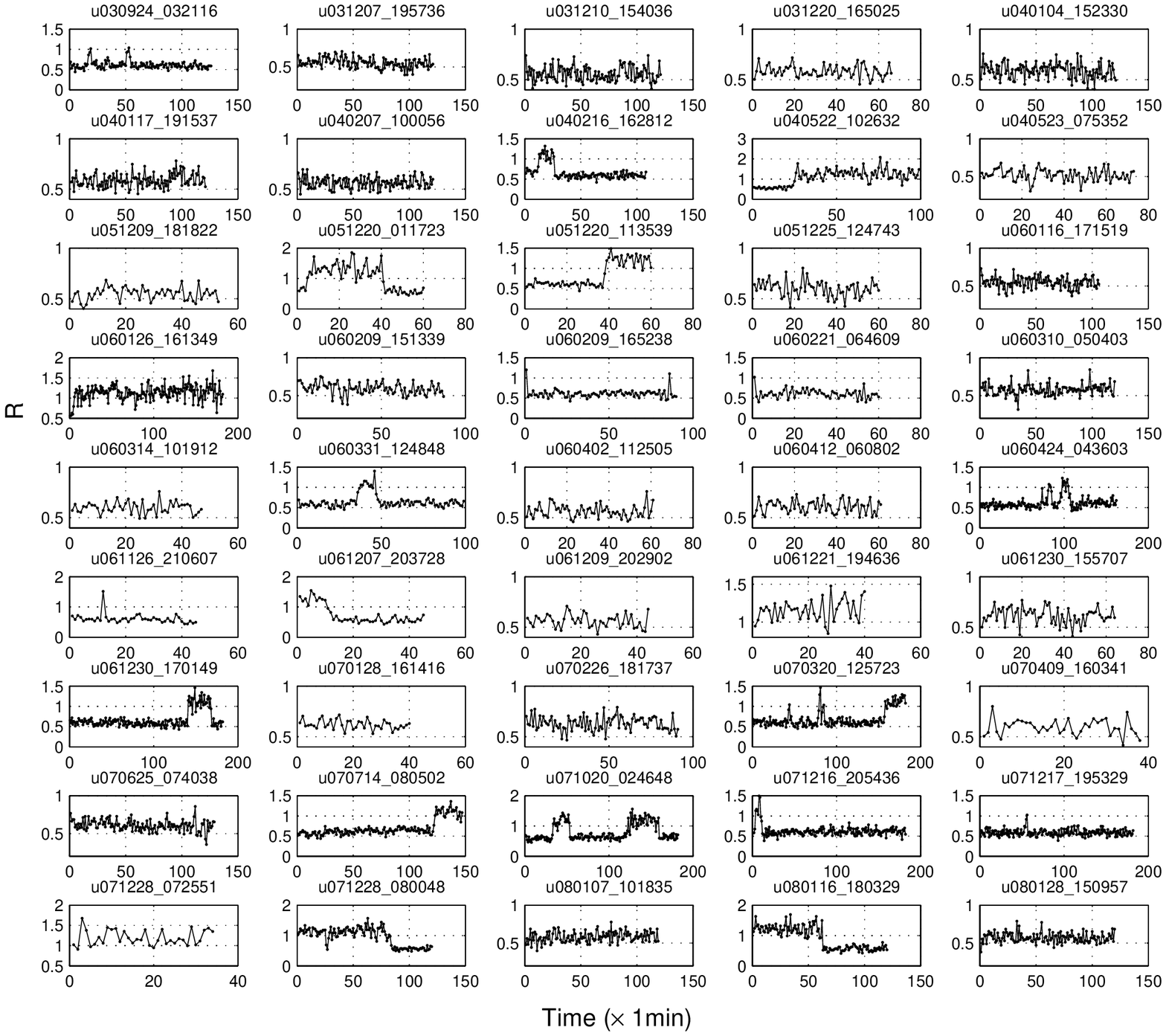}
   \caption{Time sequences of the ratio $R$ for the former 45 observations in Table \ref{table:MJD}. The integration time for the individual profiles is 1 minute.}
   \label{Fig:rseq1}
   \end{figure*}
   \begin{figure*}[h!!!]
   \includegraphics[width=18.0cm, angle=0]{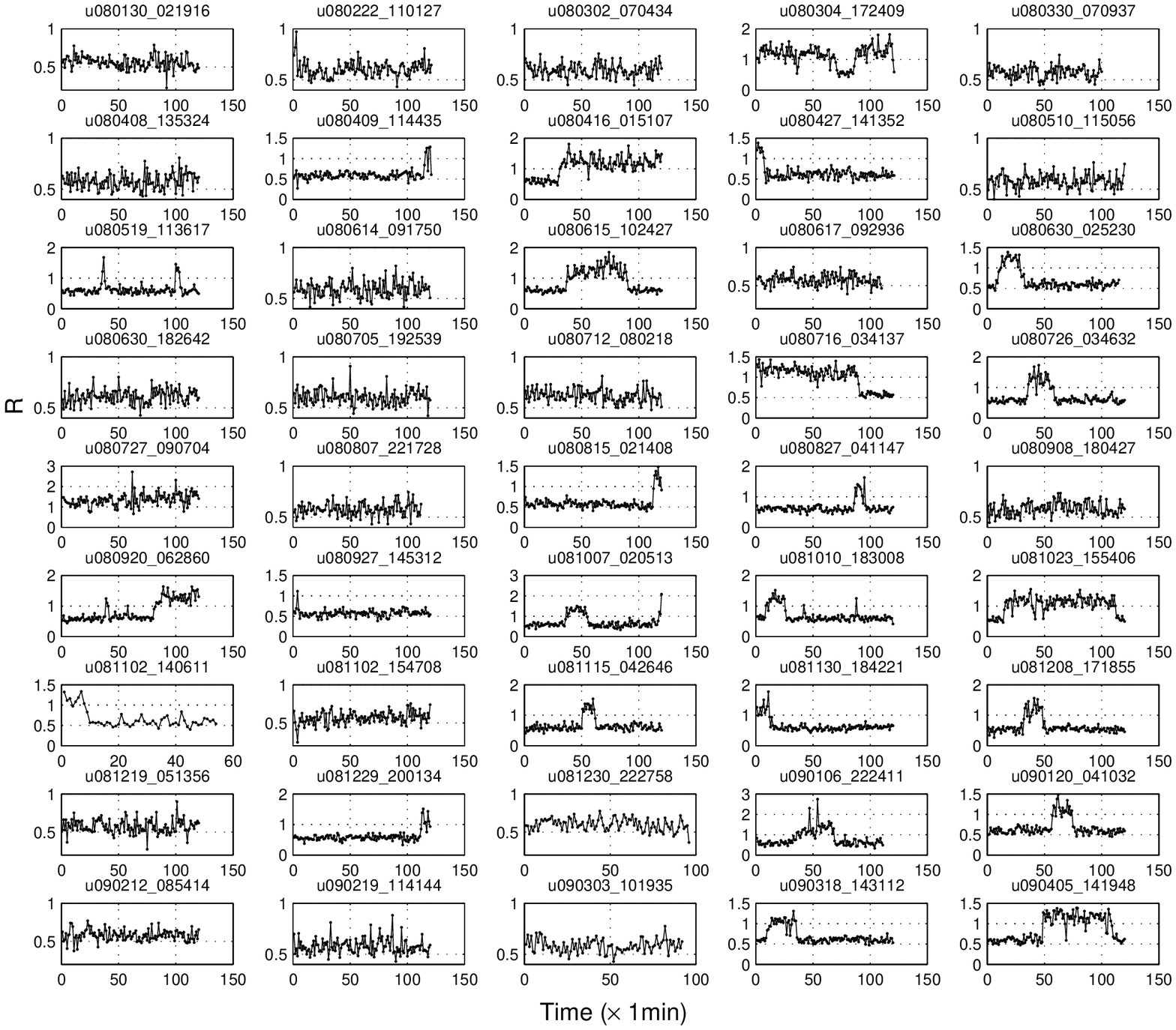}
   \caption{Time sequences of the ratio $R$ for the latter 45 observations in Table \ref{table:MJD}. The integration time for the individual profiles is 1 minute. }
   \label{Fig:rseq2}
   \end{figure*}

\section{Data reduction and Results}
\label{sect:data reduction}

At 1.5~GHz, the average pulse profile of PSR B0329+54 consists of three apparent components (Fig. \ref{Fig: pulsecomponent}). Being classified as the triple or possibly the multiple type (Rankin 1993), the integrated profiles of this pulsar can be separated into five (Kramer et al. 1994) or even nine components (Gangadhara \& Gupta 2001) by using Gaussian decomposition techniques. Whereas in this paper, we separate the profiles into three components, hereafter called component I (the leading), II (the central) and III (the
trailing), by a direct and simple way described as follows. Note that the most prominent feature of the mode changing is the variation of relative intensity between the leading and trailing components, even with our simple reduction,
the ratio $R=I_{\rm I}/I_{\rm III}$ is still a good indicator to distinguish different
modes, where $I_{\rm I}$ and  $I_{\rm III}$ represent the integrated
intensity of components I and III, respectively. The relative
intensity between components I and II, defined as $R^\prime=I_{\rm
I}/I_{\rm II}$, is also calculated, although it is not so sensitive as $R$ in identifying different modes.

\begin{figure*}[h!!!]
   \centering
   \includegraphics[width=10.0cm, angle=0]{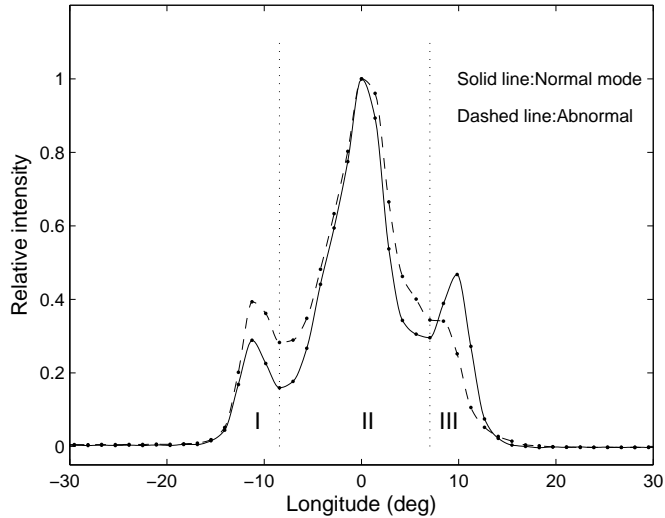}
   \caption{The average pulse profiles of PSR B0329+54 for the normal (solid)
   and abnormal modes (dashed), which are integrated over more than 2000 pulse periods respectively.
   The dots are parts of the 256 sampling points in and near the pulse window. The solid
   and dashed curves represent the smoothed profiles by cubic interpolation in a higher
   sampling resolution of $256\times$5.}
   \label{Fig: meanpulse}
   \end{figure*}

In order to calculate $R$ and $R^\prime$, the phase intervals
of components are determined by the following three steps. Firstly, a reference profile with high signal to
noise ratio was produced by summing all the profiles of one mode in an observation. Secondly, for simplicity, the outer boundaries of components I and
III are confined at 10\% level of the leading and trailing pulse
peaks of the reference profile, while the inner
boundaries are defined as the points of minimal intensity between
components I (or III) and II. Thirdly, the target profiles averaged over a few minutes are
correlated and aligned with the reference profile in the same observation. In order to reduce the uncertainty in
determining the inner boundaries, the target and reference
profiles are smoothed by cubic interpolation in a higher resolution
of 256$\times$5 sampling points per period.
The examples of reference profiles and phase boundaries are shown
in Fig \ref{Fig: meanpulse}, which are integrated over 25~min. and 15~min.
for the normal and abnormal modes, respectively.

Most previous work on mode switching of PSR B0329+54 used about 3~min
(about 250 periods) as the integration time to suppress the
pulse-to-pulse fluctuations (Lyne 1971, Bartel et al. 1982, Liu et
al. 2005). Moreover, some authors suggested that more stringently
the profile stability timescales in the normal and abnormal modes at 1.4~GHz
would be $\sim$6 ~min and $\sim$36~ min respectively (Helfand 1975).
But in this work, since a short integration time is necessary to resolve short moding timescales, we use
an integration time of 1 minute, under which the modes can be well distinguished, as demonstrated below.
\subsection{Identification of two modes}
\label{sect:discussion}
 The normal and abnormal modes are easily
distinguished according to their evident difference in $R$. The
results are presented in Figs. \ref{Fig:eightday-former} and
\ref{Fig:eightday-later} for the two 8-day quasi-continuous
observations in March 2004, and in Figs. \ref{Fig:rseq1} and
\ref{Fig:rseq2} for the remaining 90 separate observations. It shows
abrupt and simultaneous changes of $R$ and $R'$ during mode switches, despite that the
difference of $R^\prime$ between modes is less significant. In
general, the ranges of $R$ are between (0.35-0.8) for the normal mode
and (0.7-1.8) for the abnormal mode.

The number of the modes can be determined from the statistical distribution of
$R$ values. For the two apparent states separated by the abrupt changes of $R$,
by using Kolmogorov-Smirnov (KS) test, we found that the distributions of their $R$ are consistent with
the normal distribution, therefore confirmed the identification of two modes. The best fit values of the mean $R_{\mu}$ and the
standard deviation $w$ of the normal distribution at 95\% confidence level are listed in Table 2.
The probability of KS-test $p$ is also shown in the table,
where the subscript ``1'' stands for 1-minute integration. Although the probabilities
is low for the abnormal mode, they all exceed the 5\% criteria below which the null hypothesis that the data follow a
normal distribution is rejected. A KS-test for $R$ obtained with 3-minute integration time can considerably
enhance the probability of the abnormal mode, as shown by $p_3$ in the table, giving
further support to the identification of modes. More details about the histograms of the ratio $R$ with different integration time are described in the Appendix.

The $R$ distribution of abnormal mode is considerably wider than
that of the normal mode, indicating that the abnormal mode is less
stabler than the normal mode, which is consistent with the statement of Helfand (1975)
that the timescale for the abnormal mode to get stable is
longer than the normal mode.

\begin{table}
\footnotesize
\begin{center}
 \label{Tab:ratio}
\caption{The best fit parameters for the normal and abnormal modes with normal distributions.  }
  \begin{tabular}{crclccc}
  \hline
Parameters&$R_{\mu}$ & $w$ & $R^\prime_{\mu}$&$w^\prime$&$p_1$&$p_3$\\
  \hline
&Two long& quasi-continuous& observations&\\
  \hline
Normal mode& $0.546\pm0.002$& $0.082\pm0.002$& $0.127\pm0.001$& $0.027\pm0.001$ & 0.996& 0.760\\
Abnormal mode& $1.168\pm0.008$& $0.182\pm0.008$& $0.161\pm0.001$ & $0.026\pm0.001$&0.098 & 0.748\\
\hline
 &The remaining&90 observations&\\
\hline
Normal mode& $0.582\pm0.002$&$0.071\pm0.002$&$0.1445\pm0.0005$& $0.0215\pm0.0008$& 0.233  &  0.581\\
Abnormal mode& $1.176\pm0.007$&$0.170\pm0.007$& $0.1738\pm0.0013$&$0.0229\pm0.0013$ & 0.070 & 0.760\\
\hline
 \end{tabular}
\end{center}
\begin{flushleft}
$p_1$ is the probability of KS-test for the $R$ distribution with an integration time of
1 min. and $p_3$ of 3 min.
\end{flushleft}
\end{table}

\subsection{Temporal properties}
Among all the 521-hour UAO data, the total duration of the normal and
abnormal modes are about 84.9\% and 15.1\%, respectively. In the two
passages of 8-day observations, the percentage of the abnormal mode is
about 83\%. This is in agreement with the previous results (Bartel et al. 1982, Xilouris et al. 1995).

\begin{figure*}[h!!!]
   \includegraphics[width=7.0cm, angle=0]{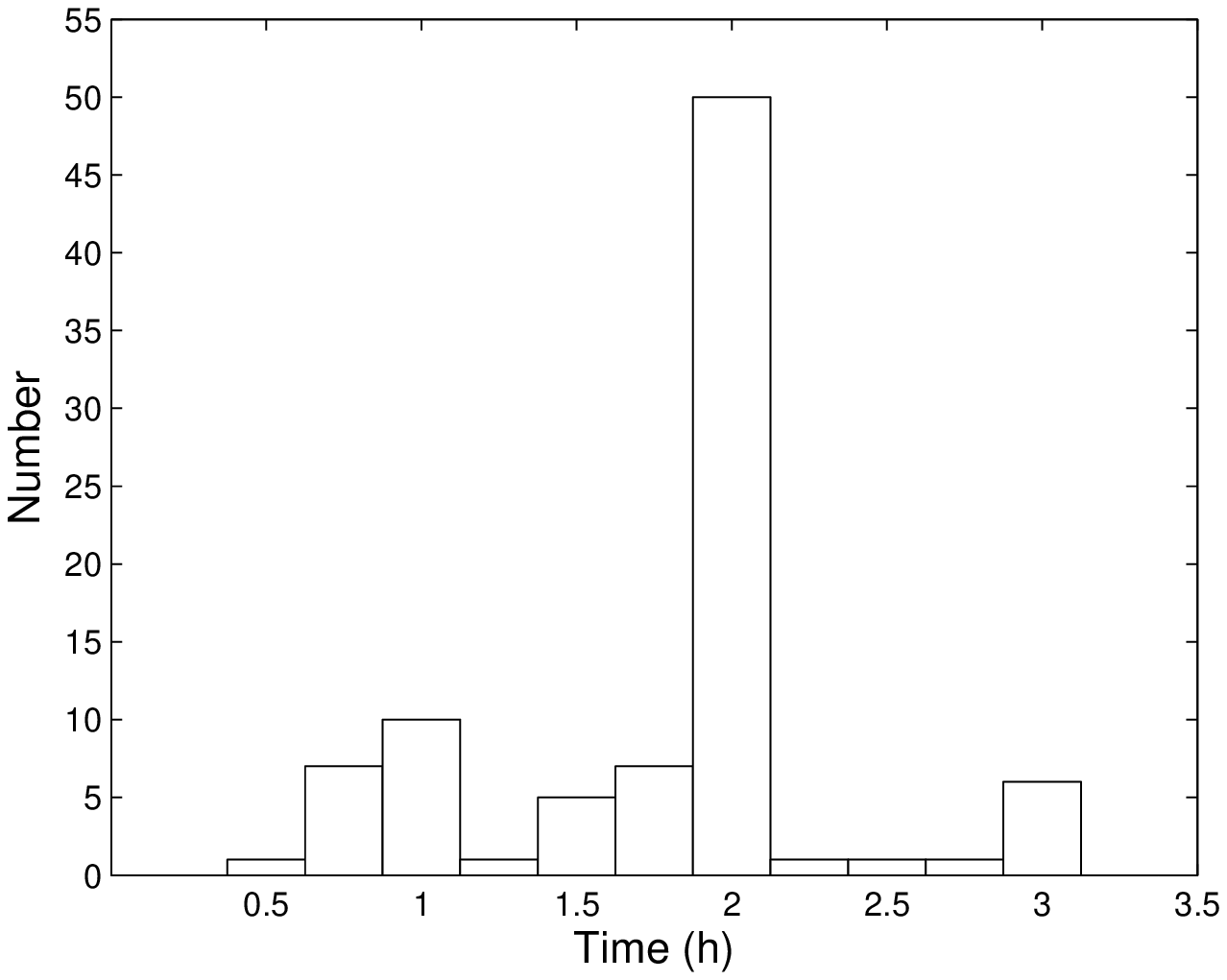}
   \includegraphics[width=7.0cm, angle=0]{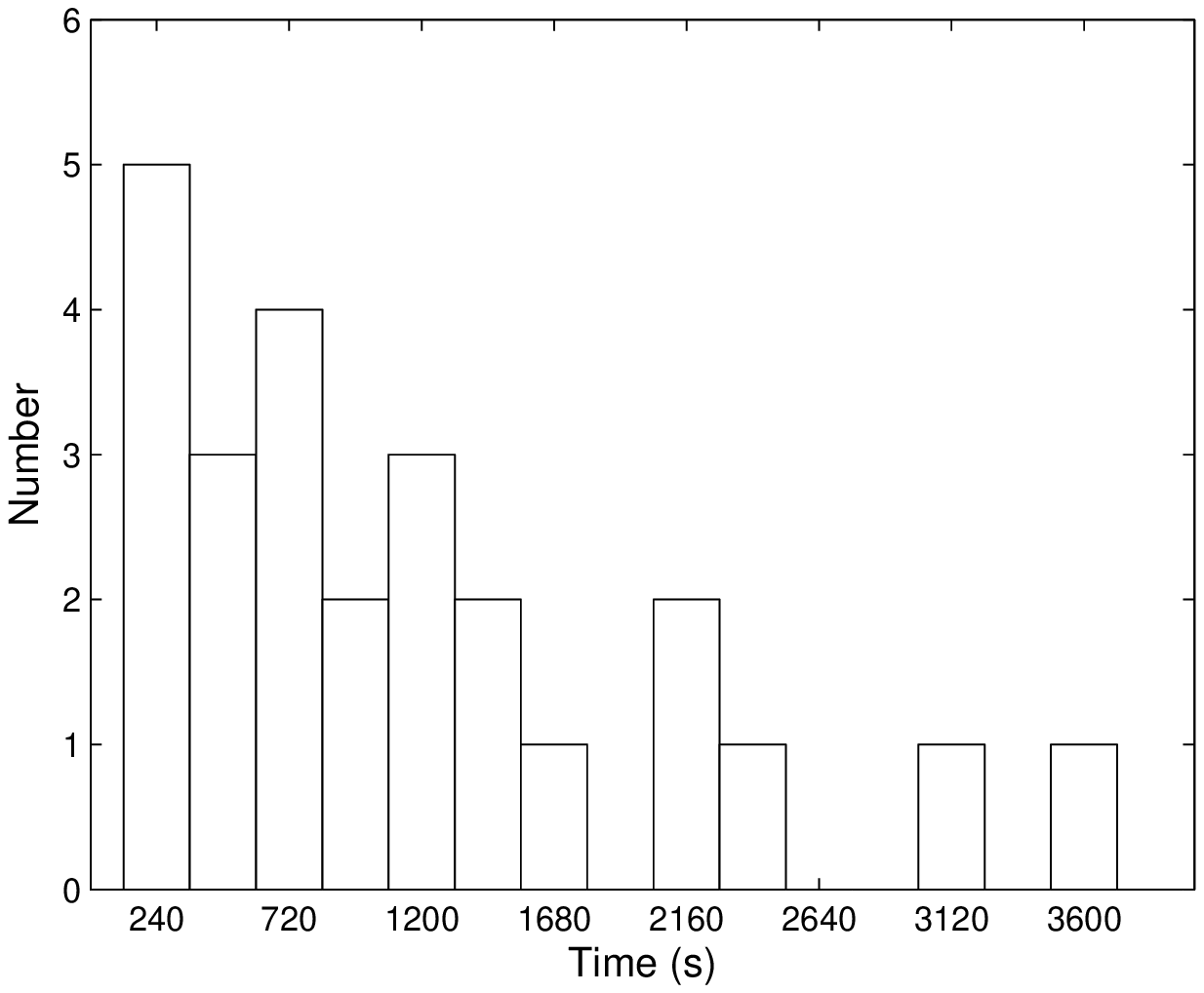}
   \caption{Left: histograms of the time duration of the 90 separated observations listed in Table \ref{table:MJD}.
   Right: histograms for the timescales of the complete sequences of the abnormal mode.}
   \label{Fig:duration}
   \end{figure*}

Totally 43 sequences of the abnormal mode are detected in the 90 separated
observations, including 25 sequences of completed abnormal
events\footnote{The completed sequence means that both the beginning and the end of a full sequence in one mode were
recorded in an observation window. The incompleted one is a sequence
truncated by the observation window, where either the intrinsic beginning or the end is missing.}. The histograms of the duration
of observation windows and the timescales of the completed abnormal
sequences are presented in the left and right panels of Fig.
\ref{Fig:duration}, respectively. It shows that the occurrence of the
abnormal mode decreases at longer timescales, but the limited
number of moding events and the strong selection effect of separated observation windows prevent us from determining meaningful results of
timescale distribution.

\begin{figure*}[h!!!]
   \centering
   \includegraphics[width=15.0cm, angle=0]{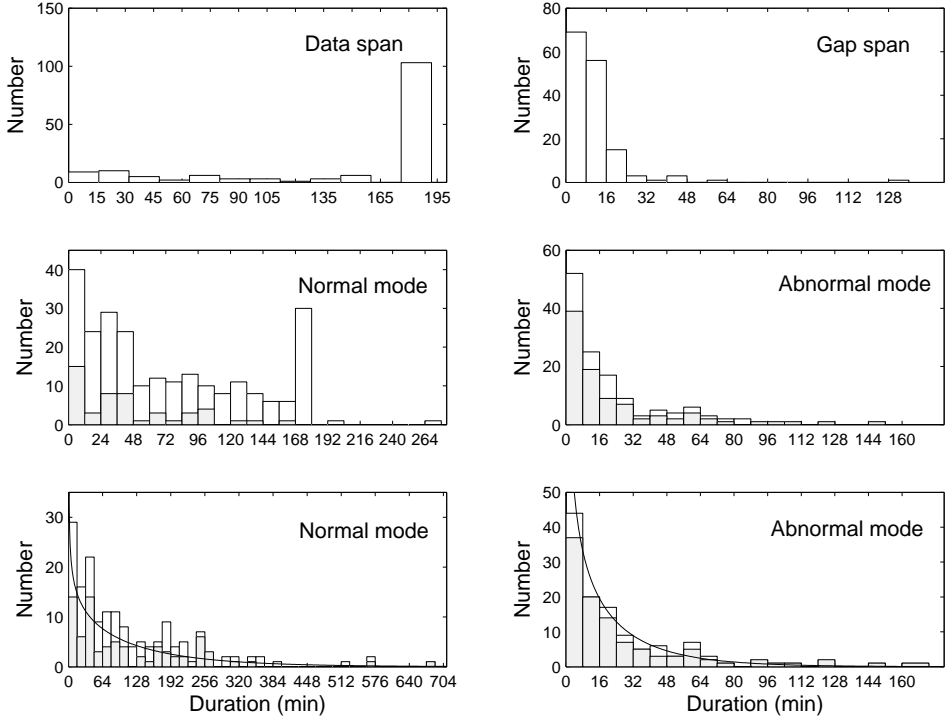}
   \caption{Upper: histograms of the time duration of the observing sub-windows from 2004 March 12 to March 31 (left) and that of the gaps between the sub-windows (right).
   Middle: histograms of the observed apparent timescales of the normal
   mode (left) and the abnormal mode (right). Lower: histograms of the corrected
   timescales of the normal mode (left) and the abnormal mode (right) after reduction
   of short gaps (see text for details). The grey bins are for the complete sequences,
   while the white bins are for both the complete and the incomplete sequences. The curves in the lower
   panels stands for the constrained optimal gamma distributions (see text for details).  }
   \label{Fig:span_8day}
   \end{figure*}

In the two long quasi-continuous observations, 244 normal and
133 abnormal sequences were identified. The two observations
consist of 148 sub-windows and 147 time gaps, of which the
histograms are shown in the upper panels of Fig. \ref{Fig:span_8day}.
Histograms made with the UAO data are shown in the middle panels of Fig.
\ref{Fig:span_8day}, where the solid bins stand for the timescales
of completed sequences, and the dashed bins for both the
completed and the incompleted sequences. Compared with the normal mode,
the proportion of the incompleted sequences of abnormal mode is much
lower. This suggests that the abnormal mode is less affected by the selection
effect of observation sub-windows, and its typical timescale is much
shorter than that of the normal mode.
%

Since the observed timescale distribution is a convolution of
intrinsic distribution and discontinuous sub-windows, it raises a
question: how can we constrain the intrinsic distribution from observations? We follow two
steps to make the constraint, first to suppress the selection effect
by inferring the states in short gaps, and then to employ Bayesian
inference to constrain the intrinsic distribution. In Bayesian
inference, the timescales of the incompleted sequences tend to bias the
constrained model parameters\footnote{Because one needs to count the
probability that the true timescale of an incomplete sequence truncated by
observing windows should be greater than the observed one, then the
likelihood of data will always favor models with flat tails of
probability density function.}, hence it is vital to fill the gaps and
increase the proportion of the completed sequences. This step is
feasible, because most time gaps in the UAO data are very short, the
states could be determined with the JBO data or be inferred reasonably
according to the trend near the gaps.

For the six gaps that are completely or partially filled by Jodrell
Bank observations, the emission states are found to be in the normal mode\footnote{This
inference is reasonable based on the fact that in all the time
intervals when the UAO and JBO observations overlapped, the modes are
the same, as shown in Figs. \ref{Fig:eightday-former} and
\ref{Fig:eightday-later}. It confirms the discovery of Bartel et al.
(1982) that modes switch simultaneously at different
frequency.}. For most of the other gaps less than 10 minutes, their modes
are inferred to be the same as those both prior and posterior to the
gaps. The probability that these assumptions are correct should be much
higher than the probability that they are wrong, because only about $1/3$ of the completed abnormal sequences
and $1/4$ of the completed normal sequences are shorter than 10 minutes, hence the emission happens to change its mode in such a short gap is low.
Furthermore, since the above mentioned six gaps in the UAO data all appear in the tracks of normal sequences and the longest one lasts for nearly 8 minutes, the confirmation of normal mode by the JBO data gives additional support for our
assumptions. The histograms for the timescales of the normal and abnormal
modes after the reduction are shown in the bottom panels of Fig.
\ref{Fig:span_8day}, where the proportion of the incompleted sequences
is reduced significantly, especially on long timescales, thus
the selection effect is considerably suppressed.

According to the Bayes' theorem, for any model $M$ with the parameter
$\theta$, the posterior probability of $\theta$ given data $D$ is
proportional to the prior probability of the parameter, $p(\theta)$,
and the probability of seeing the data given $\theta$,
$p(D|\theta)$, namely, $p(\theta|D)\propto p(\theta)p(D|\theta)$. It
gives an estimation of the probability that the parameter happens to
be valid to interpret the data. This is useful for finding the
optimal value of parameter for a model by maximizing
$p(\theta|D)$. However, even when the true model is unknown prior, one
needs to compare the Bayesian factors of some candidate models in
order to select the most likely one. The Bayesian factor $I$ is the
sum of probabilities of data given all possible parameter values in
the model $M$, i.e. $I=\int_{\theta_1}^{\theta_2}
p(D|\theta)p(\theta){\rm d}\theta$. In the case of $N$ candidate
models, if the sum of probabilities ($I$) of the other $N-1$ models
is negligible compared with the total probability of all the $N$
models, i.e. $\sum\limits_{{\rm i}=2}^{N}I_{\rm i}/\sum\limits_{{\rm
i}=1}^{N}I_{\rm i}<<1$, then the model $i=1$ is favored over the other
models (Ghosh 2007).

We select four candidate models that may fit the data, i.e. (A) the
gamma distribution with the probability density function (PDF) $p(t)=\tau^{-k}t^{k-1}{\rm e}^{-t/\tau}/\Gamma(k)$,
(B) the half normal distribution with $p(t)=2\tau^{-1}{\rm e}^{-{t^2}/{2\tau^2}}/\sqrt{2 \pi}$ ($t>0$), (C)
 the log-normal distribution with $p(t)=t^{-1}\tau^{-1}{\rm e}^{-(\ln t)^2/{2\tau^2}}/\sqrt{2\pi}$ and (D) the Pareto (power-law) distribution with $p(t)=k t^{-(k+1)}$.
Model A is a two-parameter model described with the shape $k$ and the scale
$\tau$, while B, C and D are one-parameter models. In a special case
with $k=1$, the gamma distribution reduces to the exponential distribution.
Their PDFs and statistical means are summarized in Table
\ref{Tab:Bayesian}.

Lacking knowledge of the prior probability, we assume that all the prior
probabilities are uniform within a parameter range
$\theta_1\leq\theta\leq\theta_2$, i.e.
$p(\theta)=1/(\theta_2-\theta_1)$. The maximum $\theta_2$ is set to be large enough so
that the accumulated probability at $\theta>\theta_2$ is
negligible. In practice, we use the criterion
$\int_{\theta_2}^\infty p(D|\theta){\rm
d}\theta=0.001\int_{\theta_1}^\infty p(D|\theta){\rm d}\theta$ to
estimate $\theta_2$. The second term of the Bayesian factor, i.e., the
likelihood $p(D|\theta)$, is the product of the independent
probabilities that the observed timescales $t_{\rm i}$ are
interpreted given $\theta$. For a timescale of a completed
sequence, its independent probability $p_{\rm c}(t_{\rm i}|\theta)$
is the value through substituting $\theta$ and the timescale into the PDF
$p(t)$, thus the likelihood reads
$p(D|\theta)=\prod\limits_{i=1}^{N_{\rm c}}p_{\rm c}(t_{\rm
i}|\theta)$.

\begin{figure*}[h!!!]
   \includegraphics[width=14.0cm, angle=0]{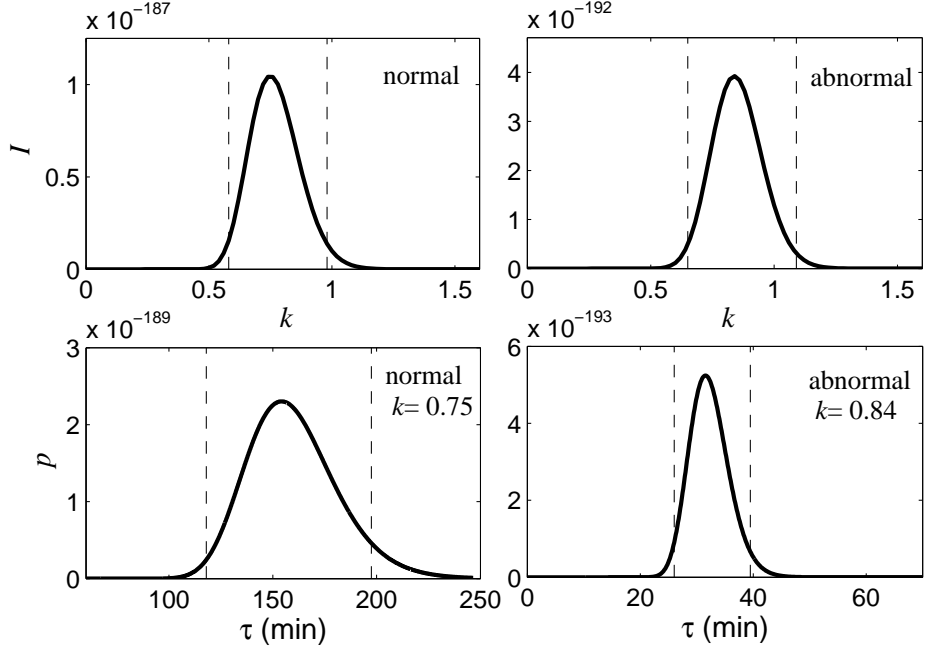}
   \caption{The Bayesian factor $I$ versus the shape parameter $k$ of the gamma distribution and the
   likelihood $p$ versus the time scale $\tau$ for the normal and abnormal modes. In the lower two panels,
   the likelihood is calculated given the optimal shape parameter $k_{\rm n}=0.75$ for the normal mode
   and $k_{\rm a}=0.84$ for the abnormal mode. The dashed lines denote the 95\% confidence intervals of $k$
   and $\tau$ around their optima.}
   \label{Fig:pdf}
   \end{figure*}

The Bayesian factors are calculated for the normal and abnormal modes
with their timescale data of the completed sequences, which are denoted
as $I_{\rm n}$ and $I_{\rm a}$ separately in Tab.
\ref{Tab:Bayesian}. The lower limit $\theta_1$, specifically to say,
$k_1$ for the Pareto and $\tau_1$ for the other three models, are set to
be 0, while the upper limits $k_2$ and $\tau_2$ are calculated following
the above criterion. As to the gamma distribution, the Bayesian factor
is calculated by integrating over the scale $\tau$, hence it is a
function of the shape $k$, i.e. $I(k)=\int_{\tau_1}^{\tau_2}
p(D|k,\tau)p(\tau){\rm d}\tau$, as shown by the upper panels in Fig. \ref{Fig:pdf}.
The optima and 95\% confidence intervals are $k_{\rm
n}=0.75^{+0.22}_{-0.17}$ for the normal mode and $k_{\rm
a}=0.84^{+0.28}_{-0.22}$ for the abnormal mode. The results are summarized from rows 3 to 8 in Tab. \ref{Tab:Bayesian}.

Among the four models, the gamma distribution is favored because its
Bayesian factors are higher than those
of the others by more than 10 orders of magnitude. Given the optimal shape parameter $k_{\rm n}=0.75$
and $k_{\rm a}=0.84$, the scale parameter is constrained to be
$\tau_{\rm n}=154^{+41}_{-36}$ minutes and $\tau_{\rm
a}=31.5^{+8.0}_{-5.5}$ minutes, respectively, through maximizing the
likelihood $p(D|k,\tau)$ and calculating the 95\%
confidence intervals (see the lower panels in Fig. \ref{Fig:pdf}). Then, the PDFs of the optimal
distributions are $p_{\rm n}(t)=0.019t^{-0.22}{\rm e}^{-t/154}$ for the
normal timescales and $p_{\rm a}(t)=0.049t^{-0.16}{\rm e}^{-t/31.5}$
for the abnormal timescales. To compare with the data, we also plot the number density functions of these two distributions in Fig. \ref{Fig:span_8day} (the curves in the lower panels). The functions reads ${\rm d}n/{\rm d}t=n_0 p(t)$, where ${\rm d}n$ is the number of the moding events within the duration $t\pm{\rm d}t$, and $p(t)$ is the PDF of gamma distribution. The constant $n_0$ is determined by minimizing $\chi^2=\sum\limits_{{\rm i}=1}^{{\rm n_b}}(N_{\rm i}-n_{\rm i})^2/n_{\rm i}$, where $N_{\rm i}$ and $n_{\rm i}$ are the observed and modeled numbers of the complete sequences in the i$-$th bin, n$_{\rm b}$ is the total number of bins. Because $N_{\rm i}$ and $n_{\rm i}$ do not follow normal distributions, this estimation is an approximation to the real number density function when n$_{\rm b}>>1$ (Press et al. 1992), and here it is only meaningful for visualization of the constrained gamma distributions and comparison with the data.

\begin{table}
\begin{center}
\caption{Results of Bayesian inference for four candidate models. }
\label{Tab:Bayesian}
\begin{tabular}{lcccc}
\hline \noalign{\smallskip}
 models  & gamma         & half normal         &     log-normal        & Pareto       \\
\hline \noalign{\smallskip}
PDF      &  $\frac{1}{\tau^k\Gamma(k)}t^{k-1}{\rm e}^{-t/\tau}$ &
$\frac{2}{\sqrt{2 \pi}\tau}{\rm e}^{-\frac{t^2}{2\tau^2}}$
                              &   $\frac{1}{\sqrt{2\pi}t\tau}{\rm e}^{-\frac{(\ln t)^2}{2\tau^2}}$
                                                      &  $\frac{k}{t^{k+1}}$     \\
mean    &  $k\tau$  & $\sqrt{\frac{2}{\pi}}\tau$ & ${\rm e}^{\frac{\tau^2}{2}}$  &   $\frac{k}{k-1}$    \\
\hline
   &  & Normal mode      &    &  \\
\hline
$\theta_{2,{\rm n}}\tablenotemark{\ast}$  & 216.5     &  199.7    & 5.69   &  0.30   \\
$\lg I_{\rm n}$     & -188.7    & -204.7    & -231.7 & -207.9   \\
($k_{\rm n}$, $\tau_{\rm n})$   & ($0.75^{+0.22}_{-0.17}, 154^{+41}_{-36}$)      & $-$      & $-$    & $-$ \\
\hline
   &  & Abnormal mode      &    &  \\
\hline
$\theta_{2,{\rm a}}\tablenotemark{\ast}$  & 45.5      & 53.8     & 3.59   &  0.50   \\
$\lg I_{\rm a}$     & -192.1    & -205.8   & -231.9 & -209.9  \\
($k_{\rm a}$, $\tau_{\rm a})$   & ($0.84^{+0.28}_{-0.22}, 31.5^{+8.0}_{-5.5}$)      & $-$      & $-$    & $-$ \\
\hline
 \end{tabular}
\begin{flushleft}
 $^{*}$Stands for $k_2$ for the Pareto distribution and $\tau_2$ for the other three (in unit of minute). $\tau$ in unit of minute.
\end{flushleft}
 \end{center}
\end{table}

\subsection{Long-term modulation}

\begin{figure*}[h!!!]
   \centering
   \includegraphics[width=11.0cm, angle=0]{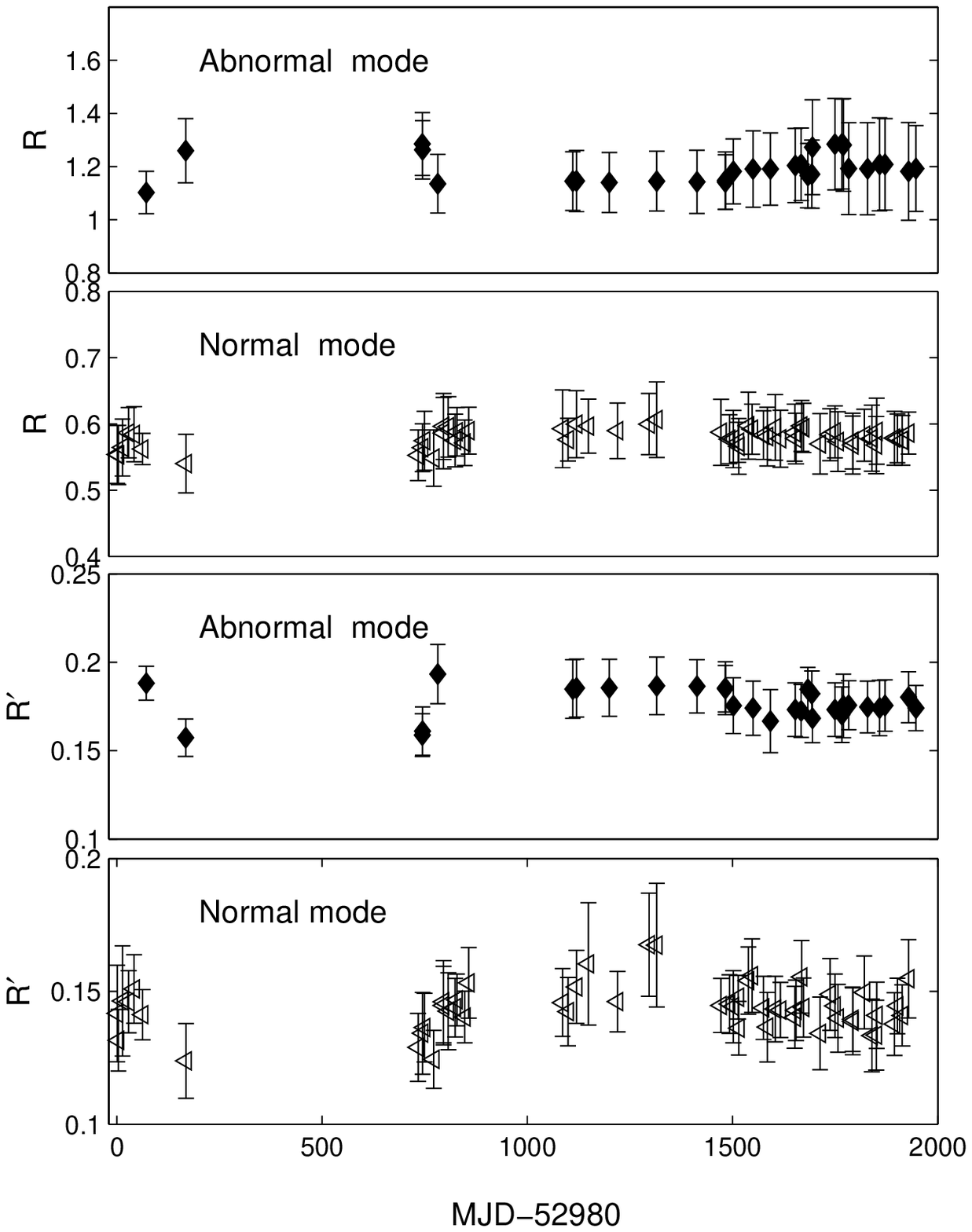}
   \caption{Time-dependent $R$ and $R^\prime$ for the normal and abnormal modes from 2003 to 2009 obtained with the UAO data. The open triangles and the solid diamonds correspond to the normal and abnormal modes, respectively. Only the observations in which the normal mode or the
abnormal mode lasts long enough are selected to make this plot.}
   \label{Fig:long-termr-normal}
   \end{figure*}

For the whole UAO observations in seven years, the ratios $R$ and
$R'$ and their $1\sigma$ errors are plotted in Fig.
\ref{Fig:long-termr-normal} for the normal (open triangles) and abnormal
(solid diamonds) modes separately. Each point stands for the
averaged $R$ for a normal or a abnormal sequence longer than 25 minutes.
The averaged ratio and the error are measured from a number of
profiles integrated over 3 minutes in that observation.

\begin{figure*}[h!!!]
   \centering
   \includegraphics[width=14.0cm, angle=0]{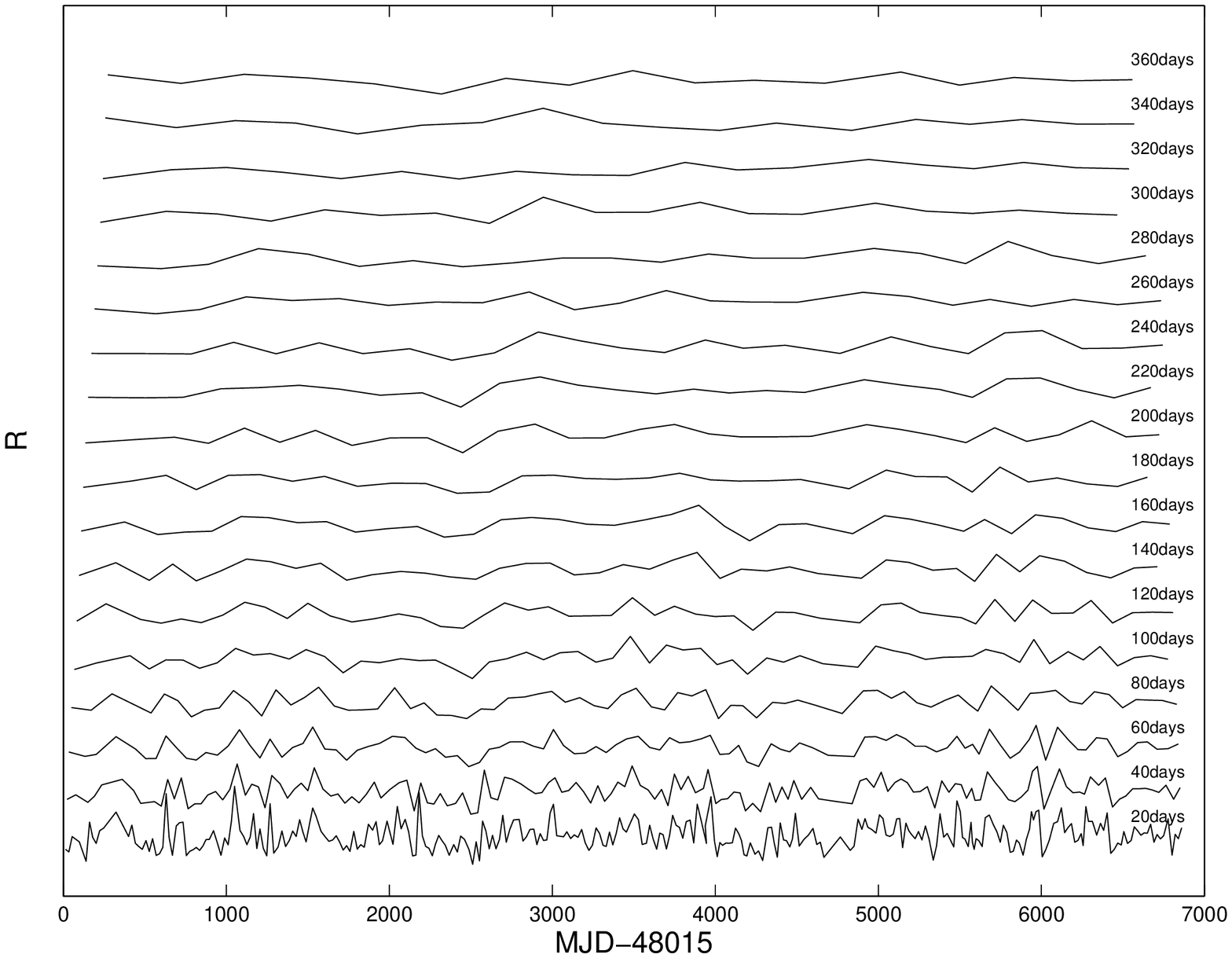}
   \caption{The long-term fluctuation of $R$ with the JBO data at 610 MHz. The numbers at right side represent the intervals in which
daily profiles are summed.}
   \label{Fig:ratio_JB}
   \end{figure*}

The JBO daily data contain 6243 useful profiles over 19
years. Fig.
\ref{Fig:ratio_JB} shows fluctuation of ratio $R$ when averaging the
profiles over different time intervals. For larger intervals, two modes tend to be
fully mixed and the ratio becomes stable. The results do not show
evident long-term variation.

\section{Conclusions and discussions}
PSR B0329+54 presents two modes based on the distribution of $R$ in
our work. Totally 176 mode switching events were detected, among
which 133 occurred during the observation in March 2004. The major
difference between the normal and abnormal modes is the intensity
ratio between the leading and trailing components and the length of moding timescales. In particular, the two passages of eight-day
quasi-continuous observations from 2004 March 12 to March 31
provide an opportunity to constrain the intrinsic distributions of
timescales for the first time. The main observational properties and
concluding remarks are summarized as follows.

(1) The intensity ratios ($R$) between the leading and trailing
components of average pulse profiles follow normal distributions, of
which the means are 0.56 for the normal mode and 1.17 for the abnormal mode,
respectively. The distribution of the abnormal mode is wider than that
of the normal mode, suggesting that it is less stabler than the normal
mode.

(2) The pulsar spent 84.9\% of the total observation time in the
normal mode, and 15.1\% in the abnormal mode. The observed longest
normal and abnormal pulse tracks lasted 4.4 and 2.47 hours,
respectively, while they are likely as long as 11 and 2.88 hours
when the mode in short gaps between observational sub-windows were
carefully inferred. No periodicity was found in the
quasi-continuous time sequences of mode switching in 2004.

(3) By virtue of the Bayesian inference, it is found that the gamma
distribution is a better description for the intrinsic distribution
of moding timescales than the other considered models, e.g. the normal,
log-normal and Pareto distributions. The optimal gamma distributions
have the shape parameter 0.75$^{+0.22}_{-0.17}$ and the scale
parameter $154^{+41}_{-36}$ minutes for the normal mode, and
$0.84^{+0.28}_{-0.22}$ and $31.5^{+8.0}_{-5.5}$
minutes for the abnormal mode.

(4) In the UAO 7-year data and the JBO 19-year data, no
prominent long-term modulation on mode switching was observed.

Mode switching was suggested by many authors to be the phenomenon that pulsars switches between different magnetospheric states, likely to be caused by changes of magnetospheric particle current flow (e.g. Lyne 1971, Bartel et al. 1982). Recently, Lyne et al. (2010) found that the changes in timing noise are directly related to the changes in pulse shape for six pulsars. The profiles of all the six pulsars switches between two states in moding timescales from a few days to hundreds of days. The change of pulsar spin-down rate $\dot{\nu}$ is highly correlated or anti-correlated with the change of pulse profiles. This discovery, together with the correlation between spin-down rate and ``on'' and ``off'' modes found in the intermittent pulsar B1931+24, give clear evidence for the mechanism of changes of magnetospheric current flow. When the current is enhanced, pulsars spin down faster and pulse profiles stay in one mode, then they switches to another mode and spin down slower when current is reduced. As to PSR B0329+54, unfortunately, the moding timescales are too short to derive useful $\dot{\nu}$. In a long period of time needed to obtain the spin-down rate, two modes are mixed in a proportion depending on the real integration time and tends to be normal/abnormal=85\%/15\% when the integration time is longer enough (see Fig. \ref{Fig:ratio_JB} as examples). Using the archived data, we did not find any correlation between the timing and the moding for B0329+54.

The underlying physical reason for the change of magnetosphere state is not clear yet. Two possible mechanisms were proposed in literature. Zhang et al. (1997) suggested that the alteration of the temperature of pulsar surface could trigger different sparking modes in the inner vacuum gap, thus results in mode switching phenomenon. It was found that the gap sparking could be dominated by magnetic absorption of the gamma-ray photons generated via curvature radiation or inverse Compton scattering (ICS). In the ICS process, the gamma-rays generated via resonant scattering of the thermal X-ray photons from the pulsar surface and non-resonant scattering of the thermal peak X-ray photons have priority in breaking down the vacuum gap. These three modes, i.e. the curvature mode, the resonant ICS mode and the thermal-peak ICS mode are separated by two critical temperatures. Thus when the temperature fluctuates around a critical value, the sparking mode will change suddenly. Although it was demonstrated that the key parameter of the ICS modes, i.e. the mean free path of the electron is a function of the temperature, the process how different sparking modes can lead to different emission modes has not been investigated. Future simultaneous and continuous radio and X-ray monitoring will be helpful to test this mechanism by searching for correlation between the emission mode and the pulsar surface temperature.

Timokhin (2010) proposed that it could be switches in the magnetosphere geometry or/and redistribution of the currents flowing in the magnetosphere that change the pulsar emission beam and its orientation with respect to the line of sight and hence lead to the mode switching phenomenon. The possible mechanism, as the author speculated, might be that the combination of current density distributions and different sizes of the co-rotating zone could result in a set of meta-stable magnetosphere configurations, which represent local minima of the total energy of magnetosphere. Some meta-stable configurations may be a kind of strange attractors, where the magnetospheric system spends a period of time in one state and then suddenly changes to another one.

Mode switching is likely to be related to stochastic dynamic processes in pulsar magnetosphere. Then, the statistical property of moding timescales is very useful to understand the physics of the non-linear system. For PSR B0329+54, as our observation shows, the waiting time until another state takes place is not quasi-periodical, but follows the gamma distributions. Further dynamic models based on, e.g. the cascade process in gaps, or the global current flow in the magnetosphere, or some sorts of instabilities, need to reproduce the observed distribution. PSR B0329+54 is yet the only pulsar with known distributions of the moding timescale. Long-term continuous monitoring for other mode-changing pulsars will find out if they have the same or various kinds of distribution, therefore will provide more constraints on the underlying physics.

\acknowledgements
We are grateful to K.J. Lee for his valuable discussion. HGW appreciate J.L. Han for his helpful comments and M. Lv for his help in data reduction. We are grateful to the referees for their valuable suggestions. This work is supported by the Knowledge Innovation Program of the Chinese Academy of Sciences, Grant No. KJCX2-YW-T09, NSFC project
10673021, 10778714 and National Basic Research Program of China (973 Program
2009CB824800), the Bureau of Education of
Guangzhou Municipality (No.11 Sui-Jiao-Ke [2009]), and GDUPS (2009).

\appendix
\label{AppendixA}

{\bf Appendix}

\begin{figure*}
 \centering
\includegraphics[width=7.0cm,angle=0]{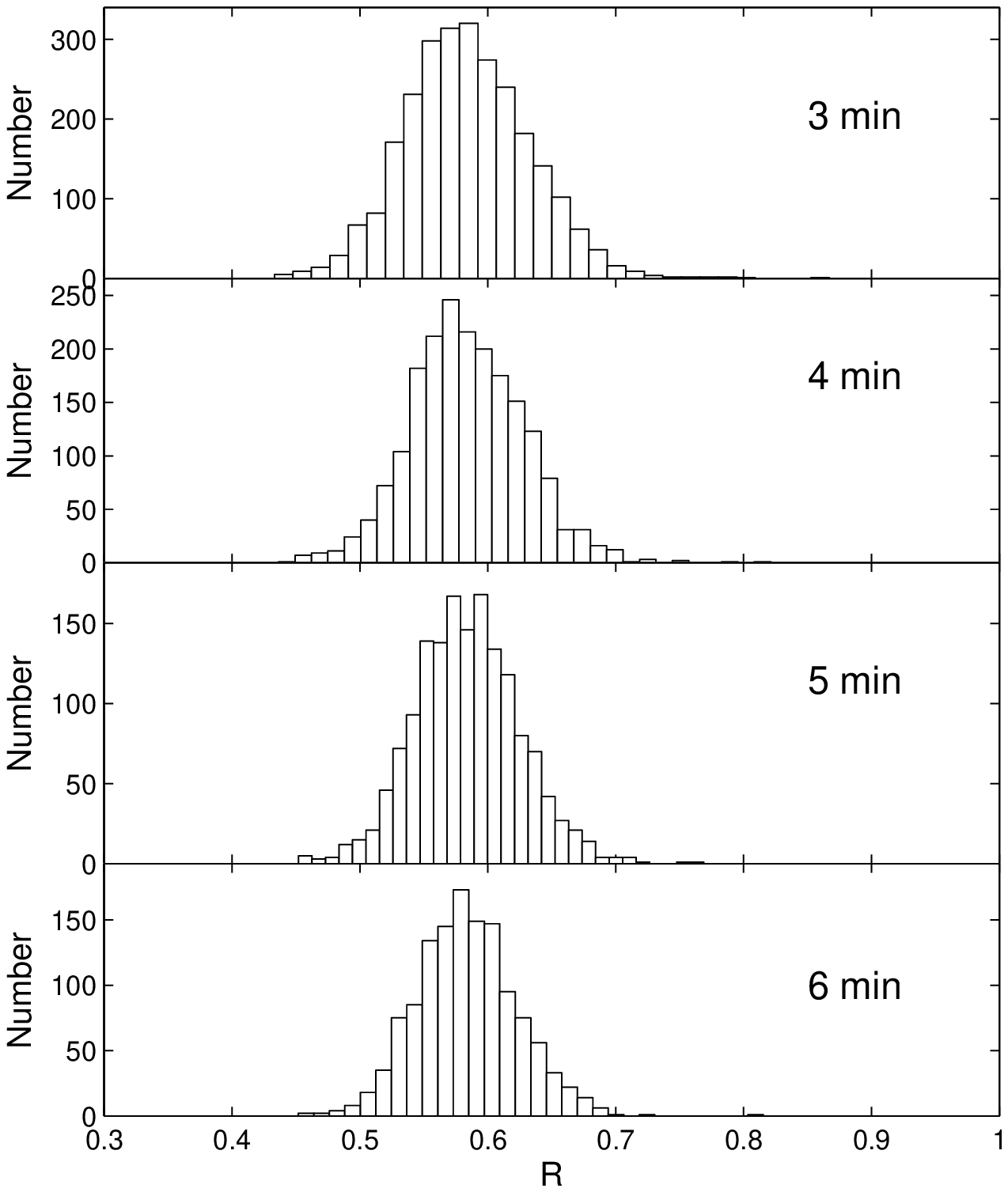}
\includegraphics[width=7.0cm,angle=0]{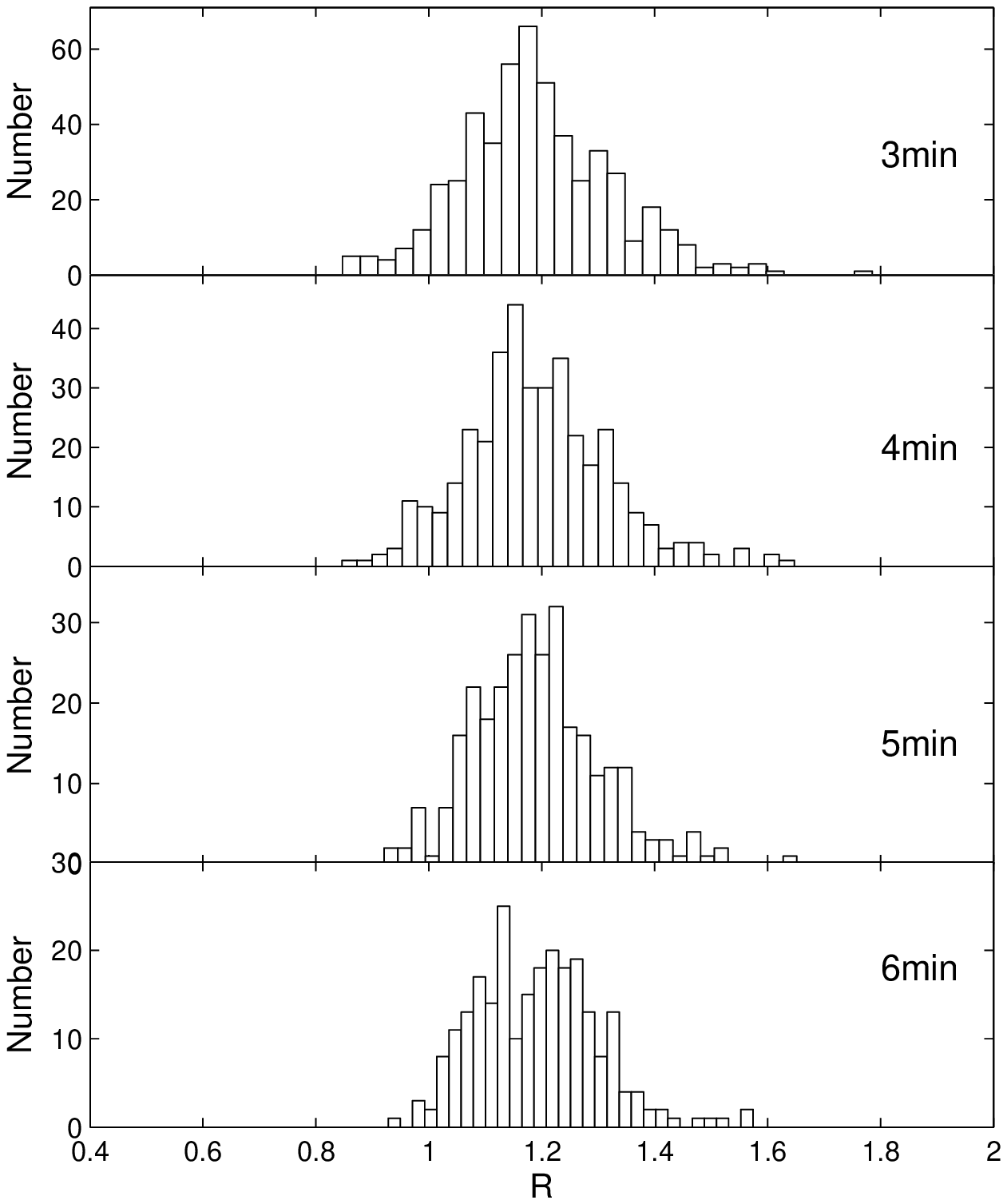}
\caption{$R$ distributions versus different integration times for the
normal (left) and abnormal (right) modes by using the data of the 90 separated observations.}
   \label{Fig: hist_4time-norm}
\end{figure*}

In order to test whether different integration times affect the
statistical properties of the intensity ratio $R$, we used 3, 4, 5 and 6
minutes as the integration time to generate the histograms of $R$, as
shown in Fig. \ref{Fig: hist_4time-norm}. The distributions appear
very similarly and can be well fitted with the normal distribution,
except that it is less scattered for large integration times. This
confirms that two modes exist in the UAO data.



\end{document}